\documentclass[a4paper,11pt]{article}
\usepackage{amssymb}
\usepackage{amsmath}
\usepackage{graphicx}
\usepackage{amsbsy}
\usepackage{xr}
\usepackage{bm}
\usepackage{color}
\usepackage{gensymb}
\usepackage{caption}
\usepackage{subcaption}
\usepackage[space]{grffile}
\definecolor{cream}{RGB}{222,217,201}
\usepackage{siunitx}
\DeclareSIUnit[number-unit-product = {\,}]
\cal{cal}
\DeclareSIUnit\kcal{\kilo\cal}
\DeclareSIUnit\kcal{\kilo\joule\per\mole}
\DeclareSIUnit\molar{\mole\per\cubic\deci\metre}
\DeclareSIUnit\Molar{\textsc{m}}
\usepackage{multirow}

\usepackage{setspace}

\usepackage[utf8]{inputenc}
\usepackage[T1]{fontenc}
\usepackage{authblk}

\usepackage[margin=0.7in]{geometry}

\begin{document}

\title{\bf Proteins -- a celebration of consilience}

\author[1,2]{Tatjana \v{S}krbi\'{c}\thanks{tskrbic@uoregon.edu}}
\author[3]{Trinh Xuan Hoang\thanks{hoang@iop.vast.ac.vn}}
\author[2,4]{Achille Giacometti\thanks{achille.giacometti@unive.it}}
\author[5]{Amos Maritan\thanks{amos.maritan@unipd.it}}
\author[1]{Jayanth R. Banavar\thanks{{\it corresponding author:} banavar@uoregon.edu}}

\affil[1]{\small \it Department of Physics and Institute for Fundamental Science, University of Oregon, Eugene, OR 97403, USA}
\affil[2]{\small \it
Dipartimento di Scienze Molecolari e Nanosistemi,
Universit\`{a} Ca' Foscari di Venezia
Campus Scientifico, Edificio Alfa,
via Torino 155, 30170 Venezia Mestre, Italy}
\affil[3]{\small \it Institute of Physics, Vietnam Academy of Science and Technology,
10 Dao Tan, Ba Dinh, Hanoi 11108, Vietnam}
\affil[4]{\small \it European Centre for Living Technologies (ECLT),
Ca' Bottacin, Dorsoduro 3911, Calle Crosera, 30123 Venezia, Italy}
\affil[5]{\small \it Dipartimento di Fisica e Astronomia,
Universit\`{a} di Padova and INFN
via Marzolo 8, 35131 Padova, Italy}

\date{}

\maketitle

\begin{abstract}
Proteins are the common constituents of {\it all} living cells. They are molecular machines that interact with each other as well as with other cell products and carry out a dizzying array of functions with distinction. These interactions follow from their native state structures and therefore understanding sequence-structure relationships is of fundamental importance. What is quite remarkable about proteins is that their understanding necessarily straddles several disciplines. The importance of geometry in defining protein native state structure, the constraints placed on protein behavior by mathematics and physics, the need for proteins to obey the laws of quantum chemistry, and the rich role of evolution and biology all come together in defining protein science. Here we review ideas from the literature and present an interdisciplinary framework that aims to marry ideas from Plato and Darwin and demonstrates an astonishing consilience between disciplines in describing proteins. We discuss the consequences of this framework on protein behavior.
\end{abstract}

\noindent {\small 18 pages; 1 table; 6 figures}\\
\noindent {\small \bf Keywords:} Helix; sheet; geometry; evolution; protein structure; interdisciplinarity.\\
\noindent {\small \bf PACS numbers:} 87.14.Ee, 87.15.-v, 87.10.+e\\

\section{Introduction}

Proteins \cite{creighton,lesk,bahar_jernigan_dill,berg} are powerful molecular machines. Small globular proteins fold into their native state structures rapidly and reproducibly and these folded forms determine their function. The folding of a globular protein is driven by hydrophobicity, the aversion of the protein backbone and some side chains to water. This causes a protein to expel the water from within its folded core thereby maximizing the self-interaction of the backbone. Furthermore, in order to enable diverse functions, one requires many distinct folded forms. This is elegantly enabled in proteins through their modular structures.

The building blocks of protein structures predicted by Pauling and his collaborators \cite{pauling_helix,pauling_sheet} and confirmed resoundingly in experiments over the decades allow for literally thousands of ways in which $\alpha$-helices and almost planar $\beta$-sheets can be assembled to yield putative native state structures of globular proteins. Form determines function and these proteins interact with each other along with other cell constituents in an orchestrated manner to enable life.

Our analysis begins with elementary mathematics. Symmetry dictates a tube as a minimalist model for our protein chain. The space-filling conformation of a discrete tube, whose axis is made up of $C^{\alpha}$ backbone atoms, is a helix with a specific pitch to radius ratio and a specific rotation angle. These two quantities along with the bond length uniquely determine all attributes of the space-filling helix including the tube radius. Remarkably, the geometries of both anti-parallel and parallel strand arrangements are predicted mathematically by considering space-filling arrangements of assemblies of zig-zag strand conformations (which are special two dimensional forms of a helix) of a tube of the same radius. The helix and the sheet, the modular building blocks of protein structures, are thus predicted by considerations of mathematics and physics. A zero parameter first principles prediction of the structures of these building blocks constitutes a re-derivation of Pauling’s classic results without invoking quantum chemistry, the planarity of the peptide bond, or the nature of the hydrogen bond.

But how are these structures realized here on earth? This is where chemistry enters the picture. We present two significant results in our paper. The first is the near perfect accord between theory and experiment. The second is the beautiful fit of the rules of quantum chemistry to the requirements of mathematics and physics. We then go on to study the complementary roles of the fields of chemistry and biology in evolution, natural selection (the proteins are the molecular targets of natural selection), and protein structure and function. Our work here celebrates the marvelous accord between seemingly distinct and highly complementary approaches to protein science from the fields of mathematics, physics, chemistry, and biology. More importantly, we hope that our work will provide a new framework and a fresh unified perspective for understanding proteins. We alert the reader that our work is primarily concerned with the geometries of the building blocks and {\it not} their assembly into the tertiary protein structure.

\section{Results and Discussion}

\subsection{Mathematics and physics}

A protein is a chain molecule which provides a natural context for its parts, an essential attribute of a machine. Thus we can distinguish between one $C^{\alpha}$ atom and another depending, not just on the identity of their side chain but, more generally,  on their sequential location, say from the $N$-terminal end. From a geometrical point of view, it is not possible to model a protein with just a single sphere (Figure 1a). One instead would need an object that is spatially extended to capture the chain topology. A sphere is a region of space carved around a point, the sphere center. A simple mathematical generalization of a sphere, resulting in an extended object, is to replace the point by a line and carve out space within a distance $\Delta$ from the line.  One then obtains a tube of radius (or thickness) $\Delta$ (Figure 1b). Helices are ubiquitous in bio-molecular structures. However, in every day life, we do not see helices often except in the context of tubes. An example is a garden hose, which is often wound into a helix. We will show below that a protein can indeed be usefully viewed as a flexible tube.

It is natural to wonder whether, instead of a tube, one might equivalently consider a chain of spheres with a railway train topology. There are at least two reasons why this is not a satisfactory alternative. First, from a symmetry perspective, a sphere looks the same when viewed from any direction. It is isotropic. However a chain is necessarily anisotropic. This is because there is a special tangent direction at any given location along a chain. An anisotropic chain comprised of isotropic spheres results in a symmetry clash. A second reason for the inappropriateness of a chain of spheres model is that it becomes an infinitesimally thin line in the continuum limit, whereas a tube is characterized by a {\it non-zero thickness}. Indeed, one might imagine the tube shown in Figure 1b to be a chain of coins rather than a chain of spheres in the continuum limit.

We use a bare-bones description of a protein and treat the axis of the protein-tube as a chain of equally spaced $C^{\alpha}$ atoms. The latter assumption is a good one because the measured bond length (over more than 4000 experimentally determined high resolution native state structures) is found to be ($3.81 \pm 0.02$)\r{A} \cite{local_sequence_structure}. The thickness of the tube can be thought of as being able to accommodate the other backbone atoms, which we do not explicitly consider in our coarse-grained approach. We will proceed by making a constructive hypothesis and assessing the consequences of this hypothesis. Our hypothesis follows from the recognition that the dominant folding mechanism of a protein is the hydrophobicity of its backbone along with the drive to maximize its self-interaction \cite{creighton,lesk,bahar_jernigan_dill,berg}, thereby attaining a space-filling folded state \cite{nature_tube}. Fortunately, for proteins, there is a wealth of experimental data accumulated over the decades, which serve to validate our hypothesis.

The rest of this section is a hopefully more accessible summary of the results of recent mathematical calculations \cite{Rose_PRE}. Remarkably, the theoretical analysis has no further assumptions, no chemistry input, and no adjustable parameters, and allows us to determine the space-filling conformations of a tube with a discrete axis with fixed bond length. We predict three secondary structures: a tightly wound space-filling helix; zig-zag strands packed into almost planar sheets, in two distinct manners corresponding to parallel and anti-parallel sheets; and hexagonal packing of straight backbone conformations, akin to a compact assembly of pencils. We do not discuss the third secondary structure in the rest of this paper because the presence of side-chains, sticking out from the backbone, results in steric clashes thereby ruling out this structure \cite{ramachandran,rose_GNR,rose_side_chains}. The steric clashes are deftly averted in both the helix and in the planar sheet conformations. These two secondary structures are in fact observed in proteins and, quite remarkably, have the same quantitative geometry as the $\alpha$-helix and two kinds of $\beta$-sheets, all elegantly predicted by Pauling and co-workers some seven decades ago \cite{pauling_helix,pauling_sheet}. What follows from assembling these modular building blocks is a library of putative native state folds, each corresponding to a distinct topological assembly through tight turns.

Figure 2 shows four sketches of a helix. The axis of a continuum helix (a helix, whose axis is continuous) (Figure 2a) necessarily spans all three dimensions and has just one character. The situation becomes more interesting, when the axis is discrete (as in a protein). There are now three distinct geometries that a helix can adopt: a generic helix rotation angle (Figure 2b) results in the $C^{\alpha}$ atoms spanning all three dimensions, as in Figure 2a; a helix rotation angle, $\varepsilon_0$, equal to $\pi$ (Figure 2c) leads to the helix axis becoming a two-dimensional zig-zag strand; and, finally, a rotation angle equal to $2\pi$ (Figure 2d) yields a one-dimensional straight line helical axis, which we will not consider further in our analysis as mentioned earlier.

Let us consider a continuum tube (the axis of the tube is continuous) of radius $\Delta$ and ask what its space-filling conformation is. We know, from our experience with a garden hose, that if we bend it too tightly (with a local radius of curvature smaller than the tube radius), we get a kink in the tube. So a tightly wound tube would have a local radius of curvature exactly equal to $\Delta$. The self-interaction of the tube is maximized by placing successive turns of the helix on top of each other and alongside each other, while ensuring that there are no self-intersections (Figures 3a and 3c). When viewed from the top, there is no empty space in the middle of the helix (Figures 3b and 3d). The tight packing fills the space within the helix and thereby maximizes the self-interaction. Mathematics teaches us that the geometry of such a space-filling helix is characterized by a universal pitch to radius ratio, $P/R$, of $2.512 \dots$ \cite{nature_tube}. This is a very helpful result but it only tells us about a dimensionless characteristic of the space-filling helix. So how we do make a more tangible (in this case, with actual lengths) connection with real proteins?

The intrinsic discrete nature of the protein backbone yields the two motifs, the helix and the strand. Furthermore, there is now the bond length of 3.81\r{A}, which will set our length scale. We take our space-filling continuum tube wound tightly with the magic dimensionless ratio $P/R=2.512\dots$ and discretize the axis with equally spaced $C^{\alpha}$ atoms. We then find mathematically \cite{Rose_PRE} that the largest rotation angle, $\varepsilon_0$, which ensures that the helix with the discrete axis remains space-filling, is approximately $99.8^{\circ}$ (Figure 4). (The procedure is very similar to that employed by Pauling and his colleagues \cite{pauling_helix,pauling_sheet}, who found the rotation angle allowing for the coherent placement of hydrogen bonds.) The bond length, the value of $P/R$, and the rotation angle completely specify ${\it all}$ characteristics of the space-filling helix. Notably, the tube radius is predicted to be $\Delta \sim 2.63$\r{A}.

Armed with these results, we now proceed to an analysis of the second building block of protein structures, strands assembled into sheets. We consider a zig-zag strand, a discretized helical conformation of a tube of radius $\Delta$ (now known through the helix analysis to be $\sim 2.63$\r{A}) with $\varepsilon_0=\pi$. A straight tube axis can be drawn through a strand in two ways -- either through the blue points (we will denote this as a blue tube) or the red points (a red tube) (Figure 2c and Figure 5). A single strand is not space-filling by itself. In order to maximize self-interactions, we need to pack strands together while ensuring that the side chains do not clash sterically. There are two distinct ways of doing this packing: the first is a blue tube alongside a blue tube (or equivalently red next to red); and the second is a blue tube next to a red tube (or equivalently red next to blue). These two packings yield distinct geometries (Figure 5). A three dimensional packing of strands is forbidden because of side-chain clashes, but the two types of assembly into planar sheets are both sterically allowed. One can construct mathematical arguments that other plausible packings, such as two or three helices twisted together or a helix alongside a strand do not fill space as efficiently as the unique space filling helix and the two kinds of sheets assembled from zig-zag strands. Interestingly, a pair of helices of opposite chiralities pack better than those with the same chirality. However, this is not a factor in protein native state structures because, as is well known experimentally, there is chiral symmetry breaking in protein $\alpha$-helices due to the left-handed nature of the constituent amino acids.

This is the essence of the theoretical analysis \cite{Rose_PRE} that enables a slew of zero-parameter predictions with the only inputs being the constructive hypothesis entailing the maximization of the self-interaction of the protein backbone along with setting the characteristic length scale through the bond length of $3.81$\r{A}.

\subsection{Chemistry provides a good fit to the dictates of mathematics}

Table 1 presents a comparison of our theoretical predictions with experimental data. For our analysis, we used 4416 structures (with complete information pertaining to all backbone atoms) from Richardsons’ Top 8000 set of high-resolution, quality-filtered protein chains (resolution $<$ 2\r{A}, 70\% PDB homology level) \cite{richardson_proteins2000_top8000}. The protein PDB codes are listed in the Supplementary Information of \v{S}krbi\'{c} et al. \cite{local_sequence_structure}. Hydrogen bonds were identified using DSSP \cite{DSSP} to extract 3595 helices, 8473 antiparallel pairs, and 4639 parallel pairs. Helices were defined to be $12$-residue segments with intra-helical hydrogen bonds ($N_iH \bullet\bullet\bullet O_{i-4}$ and $O_i\bullet\bullet\bullet HN_{i+4}$) at each residue. Antiparallel strand pairs were identified by three inter-pair hydrogen bonds at $(i,j)$, $(i+2,j-2)$, and $(i-2,j+2)$, $i \in $ strand 1, $j \in $ strand 2. To avoid possible end effects, only $(i,j)$ residue pairs were used. Parallel strand pairs were identified by four inter-pair hydrogen bonds between $(i,j-1)$, $(i,j+1)$, $(i+2,j+1)$, and $(i-2,j-1)$, $i \in $ strand 1, $j \in $ strand 2, and only the $i-th$ residue was considered. Table 1 shows the excellent accord between the mathematical predictions and experiments.

Let us begin with the $\alpha$-helix. Our central predictions \cite{Rose_PRE} are the rotation angle of around $99.8^{\circ}$, the tube radius $\Delta$ of around $2.63$\r{A}, and the pitch to radius ratio of $2.512\dots$ It is straightforward using these predictions to deduce a host of other quantities pertaining to the space-filling helix including the bond angle $\theta \sim 91.8^{\circ}$ and the dihedral angle $\mu \sim 52.4^{\circ}$ (Figures 6c and 6d and Table 1). It is interesting to note that the experimental helix is slightly squished compared to the theoretical prediction. The experimental mean value of the dihedral angle, $\mu$, is about 5\% smaller than that predicted by theory, while the mean experimental distance between the $(i,i+3)$ $C^{\alpha}$ atoms is about 3\% less than that predicted by theory.
(All these quantities nevertheless are equal to the predicted values within error estimates.) This helix squishing can be rationalized in two ways. Qualitatively, the atomic nature of the protein chain allows for space-filling to be accentuated by squeezing the helix more tightly. Quantitatively, this can arise because of the partial covalent bond character of a hydrogen bonded donor-acceptor pair allowing for a mutual distance a bit smaller than the sum of the van der Waals radii.

The two triangles $(i-1,i,i+3)$ and $(i,i+3,i+4)$ are predicted to be congruent (Figure 6b). The sides $(i-1,i)$ and $(i+3,i+4)$ are both equal to the bond length. The side $(i,i+3)$ is common to both triangles. And the angles $(i-1,i,i+3)$ and $(i,i+3,i+4)$ are both equal to $90^{\circ}$. Theory predicts that the two triangles do not lie in a plane but rather that the planes of the triangles make an angle of approximately $215.5^{\circ}$ with each other in excellent accord with the experimental data of $(213.1 \pm 5.9)^{\circ}$.

We now turn to the pairing of two strands. Unlike in a helix, the pairing of strands is necessarily non-local. As noted earlier, the chain topology of a protein naturally provides a context for the location of a $C^{\alpha}$ atom. There is a distinction between whether two paired strands are parallel to each other (meaning proceeding in the same direction) or are antiparallel to each other. As pointed out by Pauling, there are two distinct hydrogen bonding patterns for parallel and antiparallel sheets. The two distinct pairing patterns emerge from the simple tube picture as well without invoking any chemistry. The distances required by the tube constraints of being placed parallel and alongside each other (but with the correct choice of the tube axes) are very well satisfied experimentally (see Table 1). 

\subsection{Biology and chemistry}

Our discussion so far has been limited to the backbone atoms, common to all proteins, and to the secondary structures that are the modular building blocks of protein folds \cite{levitt_chothia_1976,chothia_1992,teresa,taylor_nature_2002}. So what of the amino acids and their side-chains? Amino acid sequences play a major role in facilitating amazing functionalities \cite{creighton,lesk,bahar_jernigan_dill,berg}. To illustrate a concrete example, let us consider the enzymatic function of enhancing reaction rates. The rate enhancement often arises by a lowering of the free energy of the transition state of the reaction through specific binding of the enzyme to the substrate or the reactant(s). The native state fold of the enzyme (selected from the predetermined library) has an active site, where just a few amino acids are responsible for the catalytic activity. The binding to the substrate is of course highly specific. Proteases, which are responsible for the degradation of proteins through the hydrolysis of peptide bonds, undergo convergent evolution (e.g. the digestive enzyme chymotrypsin and subtilisin, an enzyme made by soil bacteria) using distinct native state folds from the pre-sculpted library but having the same catalytic triad due to sequence design. The catalytic triad in both cases comprise three amino acids, serine, histidine, and aspartate, bound to each other by hydrogen bonds, resulting in the proton being moved away from the serine along with the creation of a reactive alkoxide ion.

Divergent evolution also occurs in proteins whose native state structure and the catalytic triad are the same, yet the nature of the binding site is not. A notable example is the family of proteins including chymotrypsin (which hydrolyses the peptide bonds on the carboxyl side of aromatic or large hydrophobic amino acids such as Trp, Tyr, Phe, Met, and Leu), trypsin (a digestive protein made in the pancreas which cleaves after positively charged amino acids lysine and arginine), elastase (made both in the pancreas and by white blood cells, which specifically targets elastin, a building block of blood vessel walls), thrombin (which cleaves proteins only at arginine-glycine linkages and helps curb bleeding by creating a blood clot), plasmin (an enzyme which cleaves proteins after lysine and arginine and dissolves blood clots), cocoonase (which also cleaves after lysine and arginine in the silk strands of the cocoon facilitating the emergence of the silk moth), and acrosin (which creates a hole in the protective sheath around the egg to allow sperm-egg contact). Key functional sites of proteins exhibit a high degree of conservation \cite{Ref_84_from_PRE_2004,konate_elife_2018} and coevolutionary analysis has been helpful in identifying protein-protein interactions \cite{Ref_85A_from_PRE_2004,Ref_85B_from_PRE_2004,Ref_86_from_PRE_2004,baker_pnas_2017}.

In accord with our findings, experiments have shown that not only can the topology of the native state be preserved on significantly changing the amino acid sequence \cite{Ref_62A_from_PRE_2004,Ref_62B_from_PRE_2004,Ref_62C_from_PRE_2004,Ref_62D_from_PRE_2004,Ref_62E_from_PRE_2004,Ref_62F_from_PRE_2004,Ref_62G_from_PRE_2004,Ref_62H_from_PRE_2004,Ref_65A_from_PRE_2004,Ref_61A_from_PRE_2004,Ref_60_from_PRE_2004,Ref_61B_from_PRE_2004,Ref_65B_from_PRE_2004,Ref_45A_from_PRE_2004,Ref_45B_from_PRE_2004,Ref_62I_from_PRE_2004,Ref_46A_from_PRE_2004,Ref_46B_from_PRE_2004,Ref_65C_from_PRE_2004,Ref_62J_from_PRE_2004}, but also the rate of protein folding does not change appreciably. Indeed, there is evidence from experiments \cite{Ref_45A_from_PRE_2004,Ref_45B_from_PRE_2004,Ref_46A_from_PRE_2004,Ref_46B_from_PRE_2004,Ref_44_from_PRE_2004,Ref_47_from_PRE_2004} that the structures of the transition states do not change much for proteins with similar native state structures. Experiments have shown that many protein sequences can have the same native state conformation \cite{Ref_64A_from_PRE_2004,Ref_64B_from_PRE_2004,Ref_64C_from_PRE_2004,Ref_64D_from_PRE_2004}. Interestingly, the same fold (e.g. the TIM (Triose phosphate IsoMerase) barrel) can facilitate multiple distinct functionalities \cite{Ref_66_from_PRE_2004,Ref_87_from_PRE_2004,franklin_elife_2018}. Also, the protein structure prediction method of threading \cite{Ref_67_from_PRE_2004} relies on the notion that a given protein does not fashion its own native state fold but rather selects from a predetermined library of folds.

Evolution, along with natural selection, allows nature to use variations on the same theme to perform myriad tasks in the living cell. Note that molecular evolution would not be able to work in this manner if protein structures changed continuously and were themselves subject to evolution. If protein structures were themselves to evolve and were also directly implicated in function, as we know they are, the structures of two interacting partners would have to co-evolve harmoniously so as not to disrupt function. Such an unlikely scenario would thwart evolution, as we know it. Our derivation of the building block geometries of protein structures from first principles provides proof that the menu of folds is in fact immutable. The mutation of a single amino acid at a time results in a random walk that forms a connected network in sequence space. However there is no similar continuous variation in structure space \cite{Ref_36_from_PRE_2004,Ref_79_from_PRE_2004}.

\section{Conclusion}

Here we have shown that, while mathematics alone is able to quantitatively predict the building blocks of protein structures, the results of chemistry and biology are remarkably consilient with the constraints placed by mathematics and physics. There have been many hints over the years that this might be true. More than eight decades ago, Bernal \cite{bernal} highlighted the common characteristics of all proteins. Some seventy years ago, Pauling \cite{pauling_helix,pauling_sheet}, the master of quantum chemistry, founded the field of molecular biology by predicting the geometries of the building blocks of protein structures using just the backbone atoms. Several years later, Ramachandran and his colleagues \cite{ramachandran} showed that the same building blocks were selected for by following quantum chemistry rules (without the need to invoke hydrogen bonds) while carefully avoiding steric clashes. As the number of protein sequences with known structure is growing enormously, the number of distinct protein folds is not. All this suggests that, even as Darwin's evolution is a major player in determining protein behavior, there is great simplicity in the protein problem, because Plato's ideas of the preeminence of geometry and symmetry are still very relevant when it comes to the library of putative native state folds.

Sequences and even functionalities evolve in order to fit within the constraints of these geometric structures, which are determined by mathematics and physics and are Platonic and not subject to Darwinian evolution. One cannot but marvel at the interplay of sequence, structure, and function of proteins and celebrate the consilience of mathematics and the sciences in shaping the field of protein science. Understanding the fit between sequence and structure in a precise mathematical manner remains an outstanding challenge.

We are delighted to dedicate this perspective to a dear friend Nitant Kenkre - an erudite scholar and a marvelous gentleman. We have all read his papers and one of us (JRB) has experienced the great pleasure of joyously interacting with Nitant. His knowledge of philosophy and mathematics and his humanity are inspiring.

\section{Materials and Methods} The PDB codes of the proteins used for our analysis are presented as Supplementary Information in \v{S}krbi\'{c} et al. \cite{local_sequence_structure}.\\

\section*{Acknowledgements}

We are indebted to George Rose for collaboration and inspiration. We are very grateful to Pete von Hippel for his warm hospitality and to him and Brian Matthews for stimulating conversations.\\

{\it Funding:} This project received funding from the European Union’s Horizon 2020 research and innovation program under the Marie Sk\l odowska-Curie Grant Agreement No 894784. The contents reflect only the authors’ view and not the views of the European Commission. Support from the University of Oregon (through a Knight Chair to JRB), International Centre of Physics at Institute of Physics, VAST under grant number ICP.2021.05 (TXH), University of Padova through ``Excellence Project 2018'' of the Cariparo foundation (AM), MIUR PRIN-COFIN2017 Soft Adaptive Networks grant 2017Z55KCW and COST action CA17139 (AG) is gratefully acknowledged. The computer calculations were performed on the Talapas cluster at the University of Oregon.\\

{\it Conflict of interest:} The authors declare that they have no conflict of interest.\\

{\it Author contributions:} JRB, AM, and T\v{S} conceived the ideas for the paper. T\v{S} carried out the calculations under the guidance of JRB. JRB wrote the paper. All authors participated in understanding the results and reviewing the manuscript.\\



\begin{figure}[htpb]
\centering
\includegraphics[width=0.6\linewidth]{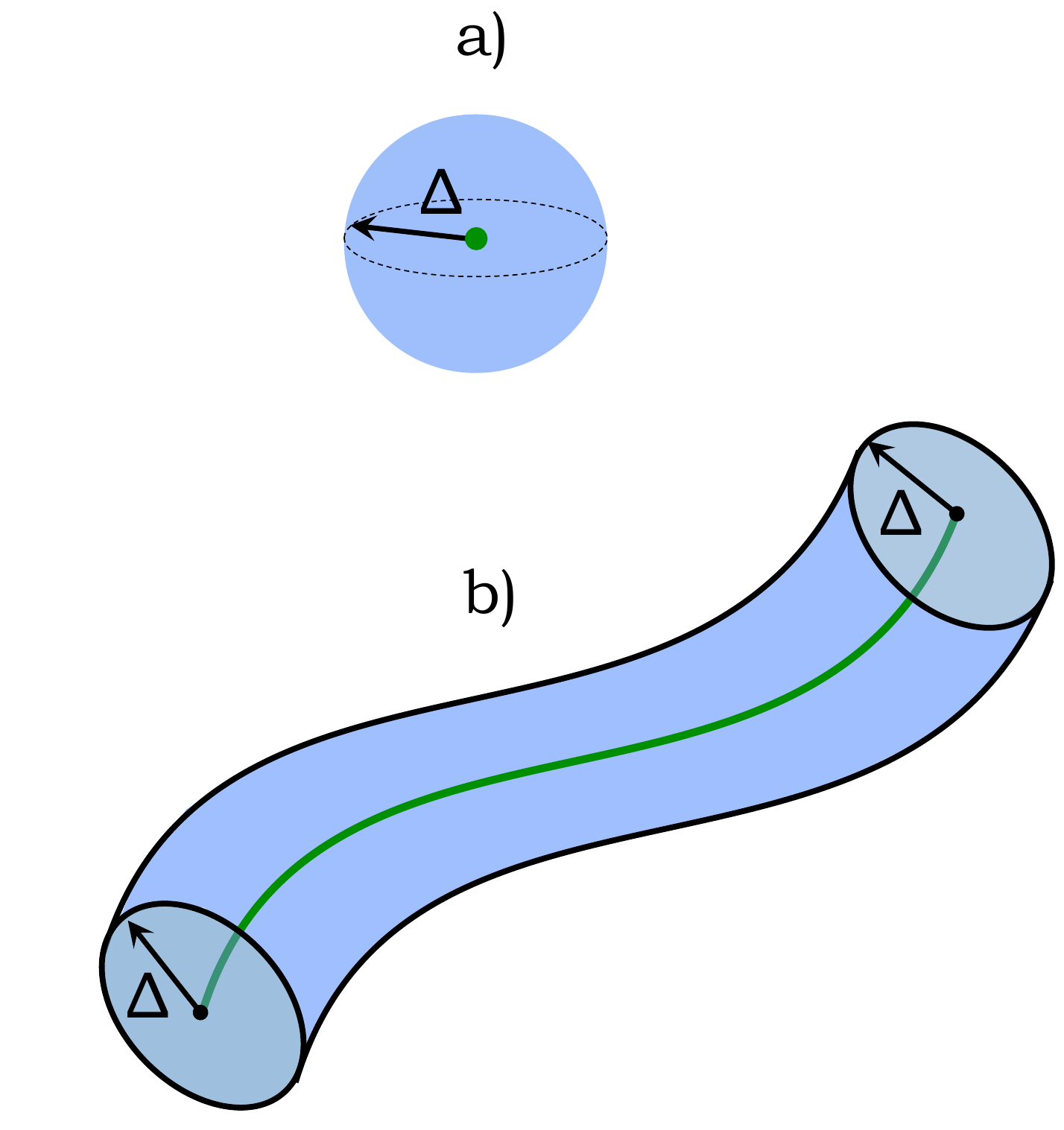}
\caption{Sketches of two simple geometries. a) A sphere of radius $\Delta$. The generalization of a sphere is b) a tube of radius $\Delta$. The green point at the center of the sphere is generalized to the green line, which is the axis of the tube. In a), the sphere encloses a region of length scale $\Delta$ around the green point, whereas, in b), the tube encloses a region of length scale $\Delta$ around the green line. A garden hose is a tube and is often curled into a helical conformation.
\label{fig:Figure_1}}
\end{figure}

\begin{figure}[htpb]
\centering
\includegraphics[width=0.8\linewidth]{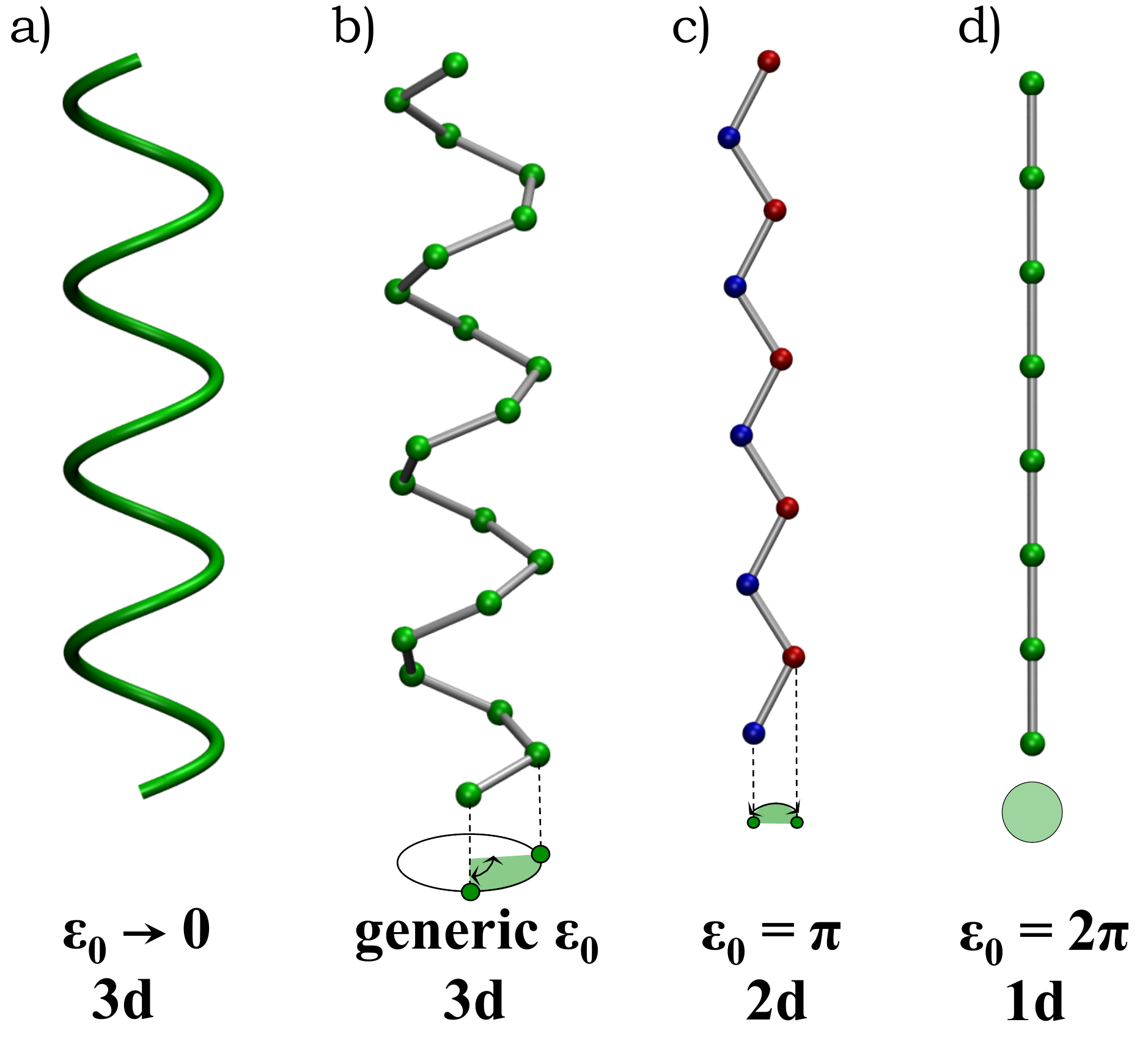}
        \caption{Sketches of four helices. a) The axis of a continuum helix with a rotation angle between successive $C^{\alpha}$ atoms, $\varepsilon_0 \rightarrow 0$, spans all three dimensions as does the axis of b) a discrete helix with a generic rotation angle $\varepsilon_0$. c) depicts the axis of a discrete helix with a rotation angle $\varepsilon_0=\pi$. The axis is a zig-zag strand spanning two dimensional space. Alternate points along the strand are colored blue and red, so that a straight line can be drawn through either the set of blue dots or the set of red dots.  d) shows the axis of a discrete helix with a rotation angle $\varepsilon_0= 2 \pi$. The helix axis is now a one dimensional straight line.}
\label{fig:Figure_2}
\end{figure}

\begin{figure}[htpb]
\centering
\includegraphics[width=0.8\linewidth]{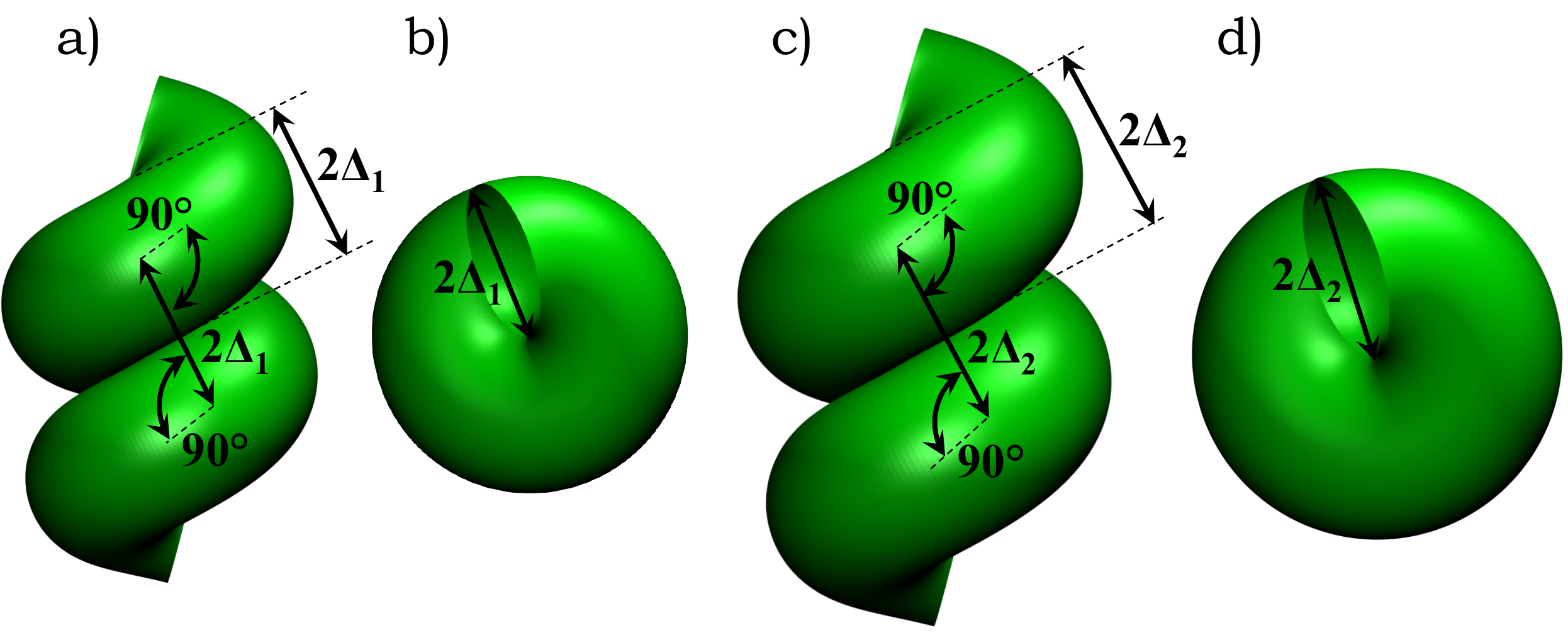}
	\caption{Sketches of two continuum space-filling helices, one with tube radius ${\Delta}_1$ and the other with tube radius ${\Delta}_2$. a) and c) show the side views of the two helices whereas b) and d) show the top views. The pitch-to-radius ratio takes on the universal value\cite{nature_tube,Rose_PRE} $P/R=2.512\dots$ for both helices. This value can be derived straightforwardly by imposing the tight-most local bending of the tube and by requiring that successive turns of the helix lie on top of each other and alongside each other. These conditions lead to the length scales shown in a) and c) being equal to the tube diameter (of the respective tubes) and all four angles shown in a) and c) being equal to $90^{\circ}$.  These geometrical conditions will be adapted for determining the characteristics of the discrete space-filling helix. Note the continuum calculation predicts the dimensionless $P/R$ ratio but has nothing to say about absolute length scales such as the tube radius $\Delta$.}
\label{fig:Figure_3}
\end{figure}

\begin{figure}[htpb]
\centering
\includegraphics[width=0.25\linewidth]{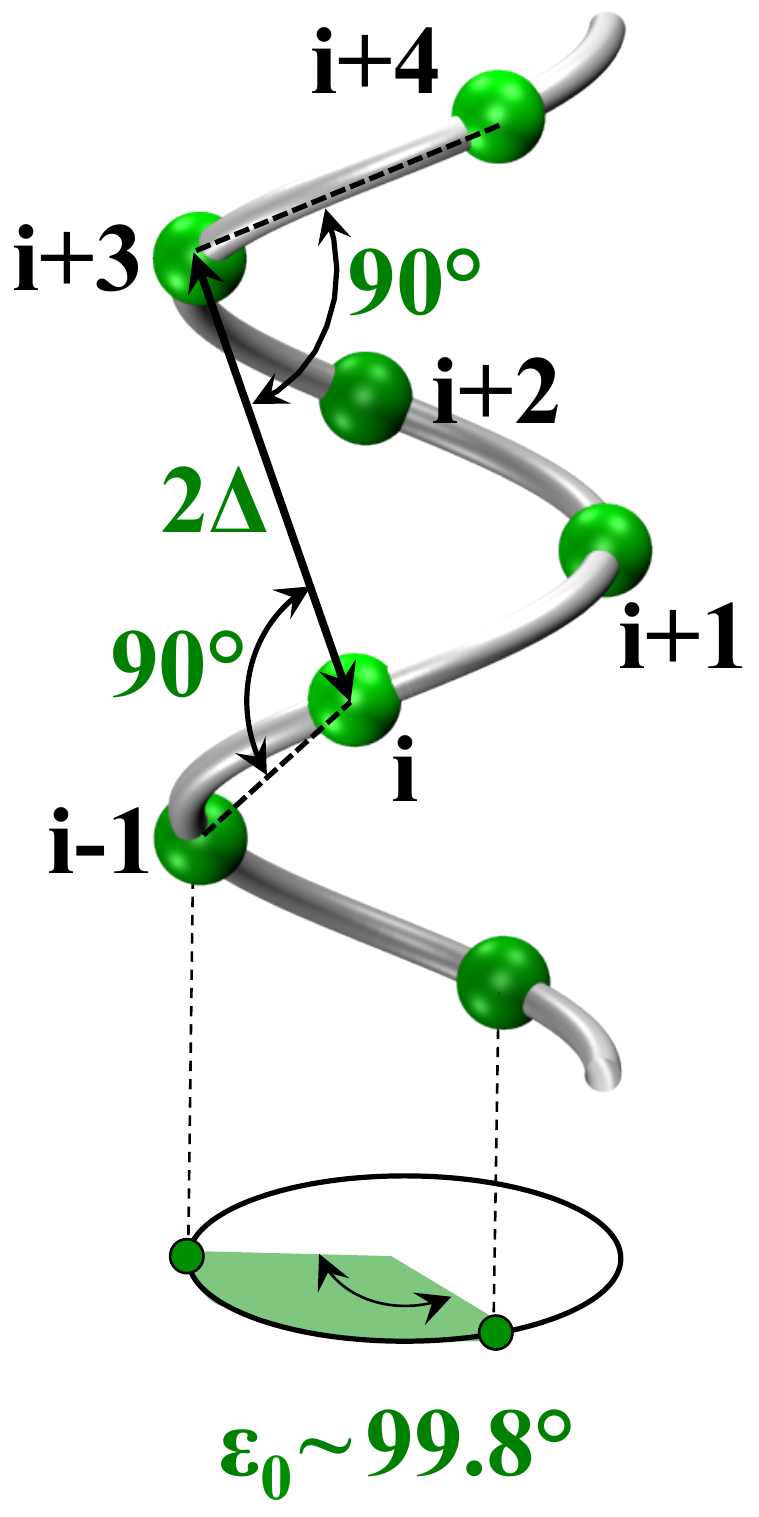}
        \caption{Sketch illustrating the derivation of the geometry of a space-filling discrete helix. The axis of a space-filling continuum helix with pitch to radius ratio $P/R=2.512\dots$ is decorated with equally spaced $C^{\alpha}$ atoms of bond length of 3.81 \r{A}, with a constant rotation angle $\varepsilon_0$. The value of $\varepsilon_0$, the bond length, and the $P/R$ ratio uniquely specify the geometry of the space-filling discrete helix and the tube radius $\Delta$. Just as Pauling determined $\varepsilon_0$ by allowing the coherent placement of hydrogen bonds, here we determine the largest $\varepsilon_0$ of around $99.8^{\circ}$ to ensure three space-filling conditions (for all $i$) adapted from the continuum calculations: the $(i,i+3)$ distance between $C^{\alpha}$ atoms is $2 \Delta$, and the two angles subtended by $(i-1,i,i+3)$ and $(i,i+3,i+4)$ $C^{\alpha}$ atoms are both equal to $90^{\circ}$.}
\label{fig:Figure_4}
\end{figure}

\begin{figure}[htpb]
\centering
\includegraphics[width=0.8\linewidth]{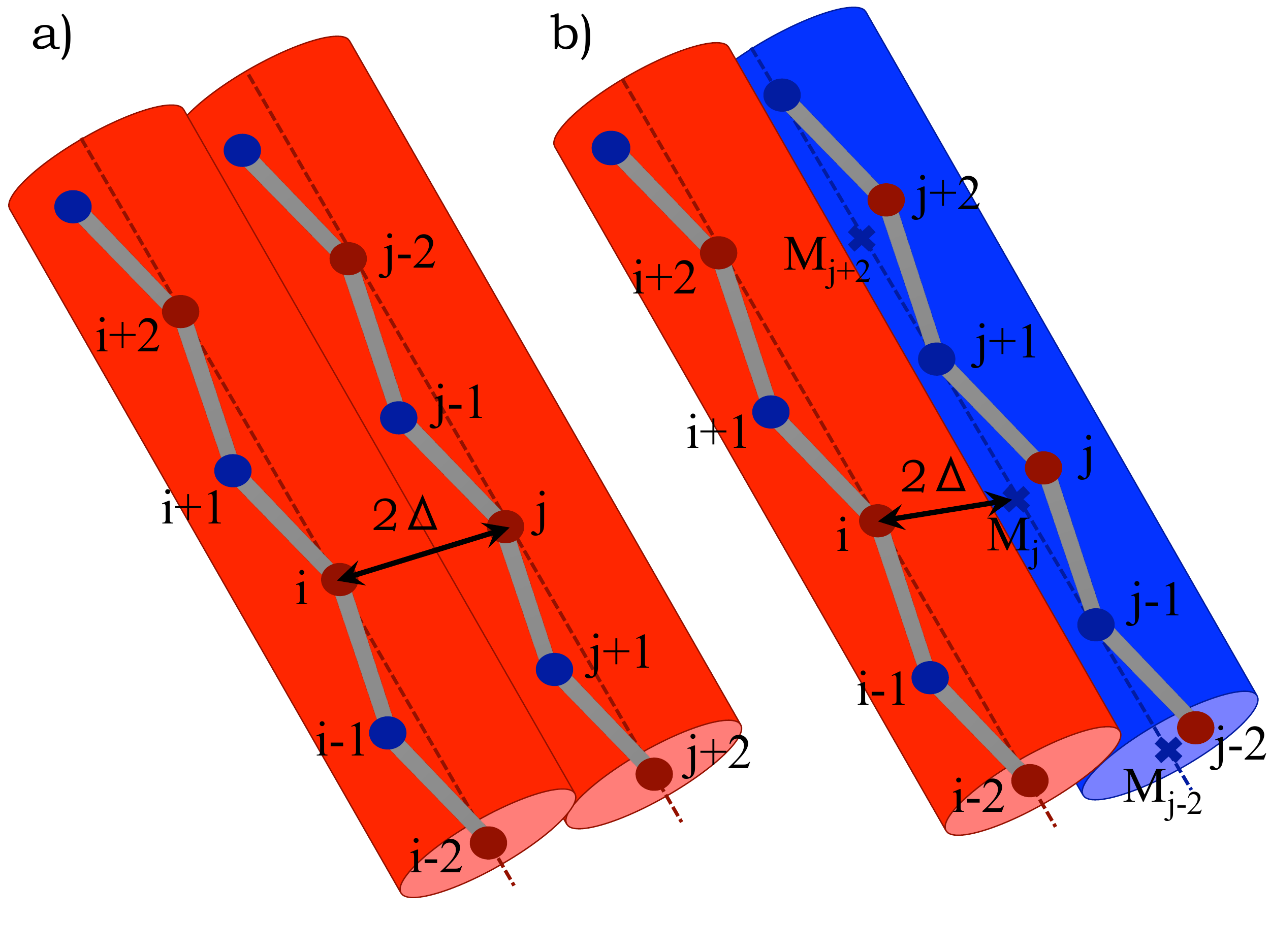}
\caption{Sketches illustrating the pairing of zig-zag strands. A zig-zag strand can be thought of as representing one of two tubes of radius $\Delta$, whose value is determined from the helical analysis. $M_j$ denotes the mid-point of sites $j-1$ and $j+1$. The two candidate tubes are the red tube whose straight line axis goes through the red points or the blue tube whose axis passes through the blue points (see Figure 2c). A space-filling pairing of strands can naturally happen in two different ways with distinct geometric constraints -- a red (blue) tube alongside a red (blue) tube or a red (blue) tube alongside a blue (red) tube. These arrangements are shown in the two panels and lead to predictions amenable to experimental validation.}
\label{fig:Figure_5}
\end{figure}

\begin{figure}[htpb]
\centering
\includegraphics[width=0.8\linewidth]{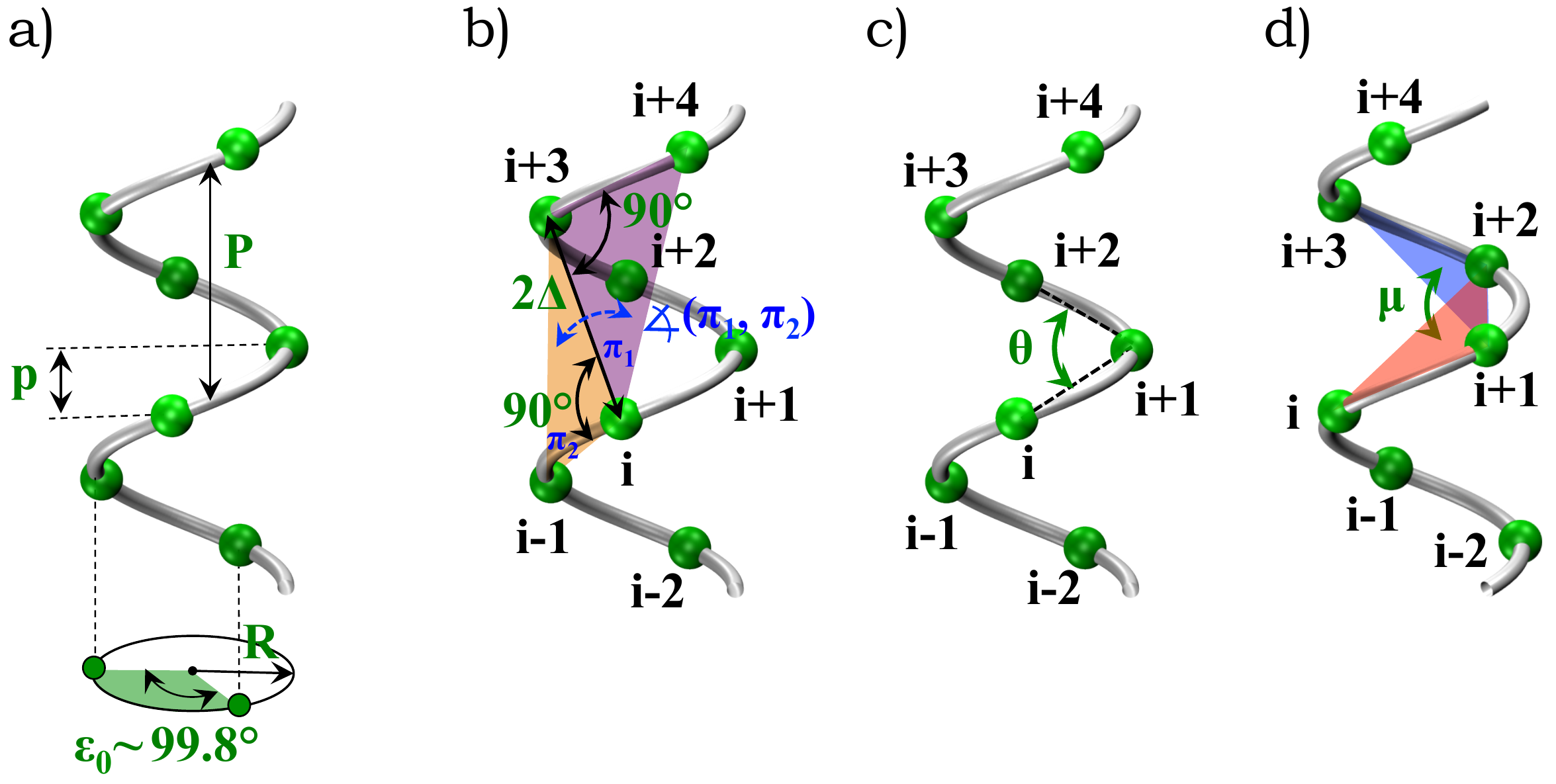}
	\caption{Sketches of quantities presented in the table for the space-filling discrete helix. a) illustrates the rotation angle $\varepsilon_0$, the rise per residue $p$, the pitch $P$, and the helix radius $R$ for the space-filling discrete helix. b) denotes a prediction pertaining to the dihedral angle between the planes defined by two congruent triangles defined by the triplets of $C^{\alpha}$ atoms $(i-1,i,i+3)$ and $(i,i+3,i+4)$. Mathematics predicts this angle to be $215.5^{\circ}$ whereas chemistry (experimental data) yields the result $(213.1 \pm 5.9)^{\circ}$. c) and d) show the definitions of the bond angle $\theta$ and the dihedral angle $\mu$ between the planes formed by two successive triplets of $C^{\alpha}$ atoms $(i,i+1,i+2)$ and $(i+1,i+2,i+3)$, respectively.}
\label{fig:Figure_6}
\end{figure}

\newpage 

\begin{table}[h]
\centering

	{\begin{tabular}{lll} 
\hline\hline         
        \multicolumn{3}{c}{\bf Continuum tube diameter from theory $2\Delta = 5.26 \dots$\r{A}}\\ [0.5ex]\hline 
{\bf Quantity} & {\bf Theory} & {\bf PDB data}\\ \hline \\ 
\multicolumn{3}{l}{{\bf HELIX}}\\[1.0ex] \hline
Rotation angle ${\varepsilon}_0$ [$^{\circ}$] & 99.8 & 99.1 $\pm$ 3.4 \\ [0.5ex] 
Number of residues per turn & 3.61 & 3.63 $\pm$ 0.13 \\ [0.5ex] 
Helix radius $R$ [\r{A}] & 2.27 & 2.30 $\pm$ 0.07 \\ [0.5ex] 
Helix pitch $P$ [\r{A}] & 5.69 & 5.47 $\pm$ 0.49 \\ [0.5ex] 
Pitch to radius ratio $c = P/R$ & 2.51 & 2.37 $\pm$ 0.29 \\ [0.5ex] 
$\angle (\pi(i-1,i,i+3), \pi(i,i+3,i+4))$ [$^{\circ}$] & 215.5 &  213.1 $\pm$ 5.9 \\ [0.5ex] 
$d(i,i+3)$ [\r{A}] & 2$\Delta$=5.26 &  5.12 $\pm$ 0.16 \\ [0.5ex] 
$\theta$ [$^{\circ}$] & 91.8 & 91.3 $\pm$ 2.2 \\ [0.5ex] 
$\mu$ [$^{\circ}$] & 52.4 & 49.7 $\pm$ 3.9 \\ [0.5ex] \hline \\ 
\multicolumn{3}{l}{\bf SHEET} \\[1.0ex] \hline
\multicolumn{3}{l}{\bf Parallel $\beta$-sheet} \\[0.5ex] \hline
$\theta$ [$^{\circ}$] & flexible & 121 $\pm$ 10 \\ [0.5ex] 
$\mu$ [$^{\circ}$] & $\sim$180 & 191 $\pm$ 17 \\ [0.5ex]
$d(i,M_j)$ [\r{A}] & 2$\Delta$=5.26 &  5.26 $\pm$ 0.16 \\ [0.5ex] 
$d(j,M_i)$ [\r{A}] & 2$\Delta$=5.26 &  4.90 $\pm$ 0.31 \\ [0.5ex] \hline 
\multicolumn{3}{l}{\bf Antiparallel $\beta$-sheet} \\[0.5ex] \hline
$\theta$ [$^{\circ}$] & flexible & 127 $\pm$ 10 \\ [0.5ex] 
$\mu$ [$^{\circ}$] & $\sim$180 & 186 $\pm$ 20 \\ [0.5ex]
$d(i,j)$ [\r{A}] & 2$\Delta$=5.26 &  5.26 $\pm$ 0.20 \\ [0.5ex] \hline \hline 
	\end{tabular}}
\label{tab:Table_1}
\label{tab:Table_1}
\caption{ {\small 
A quantitative comparison of the mathematical predictions and experimental data from protein structures.
The angle $\angle (\pi(i-1,i,i+3), \pi(i,i+3,i+4))$ is the dihedral angle between the two planes formed by the sites
$(i-1,i,i+3)$ and $(i, i+3, i+4)$, as shown in Figure 6b. $M_j$ denotes the mid-point of sites $j-1$ and $j+1$ (see Figure 5b). The anti-parallel $\beta$-sheet has a ladder-like hydrogen bond structure with a close pair of hydrogen bonds connecting symmetric sites $(i,j)$ (and $(i-2,j+2)$ and $(i+2,j-2)$). The parallel $\beta$-sheet, on the other hand, has an array of zig-zag hydrogen bonds with, for example, $i$ connected to $j-1$ and $j+1$ by a wide pair of hydrogen bonds but with $j$ not hydrogen bonded with $i+1$ or $i-1$. This breaking of symmetry between $i$ and $j$ is reflected in distinct mean experimental values of $d(i,M_j)$ and $d(j,M_i)$. The inputs to the theoretical predictions are one constructive hypothesis that the building blocks of protein structures are space-filling and the mean bond length (the average distance between adjacent $C^{\alpha}$ atoms) is 3.81\r{A}. The predictions are parameter-free and do not have any chemistry input. The quantities studied are illustrated in Figures 5 and 6. The excellent accord between theory and experiment confirm the validity of the hypothesis that self-interactions are maximized in the building blocks of protein native state structures.
	} }

\end{table}


\begin{thebibliography}{0}

\bibitem{creighton}
T. E. Creighton, {\it Proteins: Structure and Molecular Properties},
ed. W. H. Freeman (New York, 1993).

\bibitem{lesk}
A. M. Lesk, {\it Introduction to Protein Science: Architecture, Function and
Genomics} (Oxford University Press, 2~ed., 2004).

\bibitem{bahar_jernigan_dill}
I. Bahar, R. L. Jernigan and K. A. Dill, {\it Protein Actions}
(Garland Science, Taylor \& Francis Group, 2017).

\bibitem{berg}
J. M. Berg, J. L. Tymoczko, G. J. Gatto Jr. and L. Stryer,
{\it Biochemistry} (Macmillan Learning, 9~ed., 2019).


\bibitem{pauling_helix}
L. Pauling, R.~B. Corey, and H.~R. Branson,
``The structureof proteins: two hydrogen-bonded helical configurations
of the polypeptide chain'',
{\em Proc. Natl. Acad. Sci. USA} {\bf 37}, 205--210 (1951).

\bibitem{pauling_sheet}
L. Pauling, and R.~B. Corey,
``The pleated sheet, a new layer configuration of polypeptide chains'',
{\em Proc. Natl. Acad. Sci. USA} {\bf 37}, 251--256 (1951).


\bibitem{local_sequence_structure}
T.~\v{S}krbi\'{c}, A.~Maritan, A.~Giacometti and J.~R. Banavar,
``Local sequence-structure relationships in proteins'',
{\em Prot. Sci.} {\bf 30}, 818--829 (2021).

\bibitem{nature_tube}
A. Maritan, C. Micheletti, A. Trovato and J.~R. Banavar,
``Optimal shapes of compact strings'',
{\em Nature} {\bf 406}, 287--290 (2000).


\bibitem{Rose_PRE}
T.~{\v{S}}krbi{\'c}, A.~Maritan, A.~Giacometti, G.~D. Rose and J.~R. Banavar,
``Building blocks of protein structures -- Physics meets Biology'',
{\it Phys. Rev. E}, {\bf 104}, 014402 (2021).


\bibitem{ramachandran}
G.~N. Ramachandran and V. Sasisekharan,
``Conformation of polypeptides and proteins'',
{\em Adv. Prot. Chem.} {\bf 23}, 283--438 (1968).

\bibitem{rose_GNR}
G.~D.~Rose, ``In Memoriam: Professor G.N. Ramachandran (1922 -- 2001)'',
Perspective -- {\em Proteins: Structure, Function and Bioinformatics} {\bf 10},
1691--1693 (2001).

\bibitem{rose_side_chains}
G.~D.~Rose,
``Ramachandran maps for side chains in globular proteins''.
{\em Proteins: Structure, Function and Bioinformatics} {\bf 87}, 357--364 (2019).

\bibitem{richardson_proteins2000_top8000}
S.~C. Lovell, J.~M. Word, J.~S. Richardson and D.~C. Richardson,
``The penultimate rotamer library'',
{\em Proteins: Structure, Function and Bioinformatics } {\bf 40}, 389--408 (2000).
Also, see the web site: http://kinemage.biochem.duke.edu/databases/top8000.php

\bibitem{DSSP}
W. Kabsch and C. Sander,
``Dictionary of protein secondary structure: pattern recognition of hydrogen-bonded and geometrical features'',
{\em Biopolymers} {\bf 22}, 2577--2637 (1983).



\bibitem{levitt_chothia_1976}
M. Levitt and C. Chothia, ``Structural patterns in globular proteins'',
{\em Nature} {\bf 261,} 552--558 (1976).

\bibitem{chothia_1992}
C. Chothia, ``One thousand families for the molecular biologist'',
{\em Nature} {\bf 357}, 543--544 (1992).

\bibitem{teresa}
T. Przytycka, R. Aurora and G.~D. Rose,
``A protein taxonomy based on secondary structure'',
{\em Nat. Struct. Biol.} {\bf 6}, 672--682 (1999).

\bibitem{taylor_nature_2002}
W.~R. Taylor, ``A ‘periodic table’ for protein structures'',
{\em Nature} {\bf 416} 657--660 (2002).




\bibitem{Ref_84_from_PRE_2004}
O. Lichtarge and M.~E. Sowa,
``Evolutionary predictions of binding surfaces and interactions'',
{\em Curr. Opin. Struct. Biol.} {\bf 12}, 21--27 (2002).

\bibitem{konate_elife_2018}
M. M. Konaté, G. Plata, J. Park, D.~R. Usmanova, H. Wang and D. Vitkup,
``Molecular function limits divergent protein evolution on planetary timescales'',
{\em eLife} {\bf 8}, e39705 (2019).

\bibitem{Ref_85A_from_PRE_2004}
C.~S. Goh and F.~E. Cohen,
``Co-evolutionary Analysis Reveals Insights into Protein–Protein Interactions'',
{\em J. Mol. Biol.} {\bf 324}, 177--192 (2002).

\bibitem{Ref_85B_from_PRE_2004}
P.~J. Bickel, K.~J. Kechris, P.~C. Spector, G.~J. Wedemayer and A.~N. Glazer,
``Finding important sites in protein sequences'',
{\em Proc. Natl. Acad. Sci. USA} {\bf 99}, 14764--14771 (2002).

\bibitem{Ref_86_from_PRE_2004}
S.~J. Campbell, N.~D. Gold, R.~M. Jackson, and D.~R. Westhead,
``Ligand binding: functional site location, similarity and docking'',
{\em Curr. Opin. Struct. Biol.} {\bf 13}, 389--395 (2003).

\bibitem{baker_pnas_2017}
I. Anishchenko, S. Ovchinnikov, H. Kamisetty and D. Baker,
``Origins of coevolution between residues distant in protein 3D structures'',
{\em Proc. Natl. Acad. Sci. USA} {\bf 114}, 9122--9127 (2017).




\bibitem{Ref_62A_from_PRE_2004}
J.~S. Richardson and D.~C. Richardson,
``The de novo design of protein structures'',
{\em Trends Biochem. Sci.} {\bf 14}, 304--309 (1989).

\bibitem{Ref_62B_from_PRE_2004}
W.~F. DeGrado, Z.~R. Wasserman and J.~D. Lear,
``Protein design, a minimalist approach'',
{\em Science} {\bf 243}, 622--628 (1989).

\bibitem{Ref_62C_from_PRE_2004}
M.~H. Hecht, J.~S. Richardson, D.~C. Richardson, and R.~C. Ogden,
``De novo design, expression, and characterization of Felix: a four-helix bundle protein of native-like sequence'',
{\em Science} {\bf 249}, 884--891 (1990).

\bibitem{Ref_62D_from_PRE_2004}
C.~P. Hill, D.~H. Anderson, L. Wesson, W.~F. DeGrado and D. Eisenberg,
``Crystal structure of alpha 1: implications for protein design'',
{\em Science} {\bf 249}, 543--546 (1990).

\bibitem{Ref_62E_from_PRE_2004}
C. Sander and R. Schneider,
``Database of homology-derived protein structures and the structural meaning of sequence alignment'',
{\em Proteins: Structure, Function and Bioinformatics} {\bf 9}, 56--68 (1991).

\bibitem{Ref_62F_from_PRE_2004}
S. Kamtekar, J.~M. Schiffer, H.~Y. Xiong, J.~M. Babik and M.~H. Hecht,
``Protein design by binary patterning of polar and nonpolar amino acids'',
{\em Science} {\bf 262}, 1680--1685 (1993).

\bibitem{Ref_62G_from_PRE_2004}
A.~P. Brunet, E.~S. Huang, M.~E. Huffine, J.~E. Loeb, R.~J. Weltman and M.~H. Hecht,
``The role of turns in the structure of an $\alpha$-helical protein'',
{\em Nature} {\bf 364}, 355--358 (1993).

\bibitem{Ref_62H_from_PRE_2004}
A.~R. Davidson and R.~T. Sauer,
``Folded proteins occur frequently in libraries of random amino acid sequences'',
{\em Proc. Natl. Acad. Sci. USA} {\bf 91}, 2146--2150 (1994).

\bibitem{Ref_65A_from_PRE_2004}
E. Shakhnovich, V. Abkevich and O. Ptitsyn,
``Conserved residues and the mechanism of protein folding'',
{\em Nature} {\bf 379}, 96--98 (1996).

\bibitem{Ref_61A_from_PRE_2004}
D.~S. Riddle, J.~V. Santiago, S.~T. Bray-Hall, N. Doshi, V.~P. Grantcharova, Q. Yi and D. Baker,
``Functional rapidly folding proteins from simplified amino acid sequences'',
{\em Nat. Struct. Biol.} {\bf 4}, 805--809 (1997).

\bibitem{Ref_60_from_PRE_2004}
D. Perl, C. Welker, T. Schindler, K. Schr\''{o}der, M. A. Marahiel, R. Jaenicke and F. X. Schmid,
``Conservation of rapid two-state folding in mesophilic, thermophilic and hyperthermophilic cold shock proteins'',
{\em Nat. Struct. Biol.} {\bf 5}, 229--235 (1998).

\bibitem{Ref_61B_from_PRE_2004}
D.~E. Kim, H. Gu and D. Baker,
``The sequences of small proteins are not extensively optimized for rapid folding by natural selection'',
{\em Proc. Natl. Acad. Sci. USA} {\bf 95}, 4982--4986 (1998).

\bibitem{Ref_65B_from_PRE_2004}
L.~A. Mirny, V.~I. Abkevich and E.~I. Shakhnovich,
``How evolution makes proteins fold quickly'',
{\em Proc. Natl. Acad. Sci. USA} {\bf 95}, 4976--4981 (1998).

\bibitem{Ref_45A_from_PRE_2004}
V. Villegas, J.~C. Martinez, F.~X. Aviles, and L. Serrano,
``Structure of the transition state in the folding process of human procarboxypeptidase A2 activation domain'',
{\em J. Mol. Biol.} {\bf 283}, 1027--1036 (1998).

\bibitem{Ref_45B_from_PRE_2004}
F. Chiti, N. Taddei, P.~M. White, M. Bucciantini, F. Magherini, M. Stefani and C.~M. Dobson,
``Mutational analysis of acylphosphatase suggests the importance of topology and contact order in protein folding'',
{\em Nat. Struct. Biol.} {\bf 6}, 1005--1009 (1999).

\bibitem{Ref_62I_from_PRE_2004}
M.~W. West, W. Wang, J. Patterson, J.~D. Mancias, J.~R. Beasley and M.~H. Hecht,
``De novo amyloid proteins from designed combinatorial libraries'',
{\em Proc. Natl. Acad. Sci. USA} {\bf 96}, 11211--11216 (1999).

\bibitem{Ref_46A_from_PRE_2004}
J.~C. Martinez and L. Serrano,
``The folding transition state between $SH_3$ domains is conformationally restricted and evolutionarily conserved'',
{\em Nat. Struct. Biol.} {\bf 6}, 1010--1016 (1999).

\bibitem{Ref_46B_from_PRE_2004}
D.~S. Riddle, V.~P. Grantcharova, J.~V. Santiago, E. Alm, I. Ruczinski and D. Baker,
``Experiment and theory highlight role of native state topology in $SH_3$ folding'',
{\em  Nat. Struct. Biol.} {\bf 6}, 1016--1024 (1999).

\bibitem{Ref_65C_from_PRE_2004}
M. Vendruscolo, E. Paci, C.~M. Dobson and M. Karplus,
``Three key residues form a critical contact network in a protein folding transition state'',
{\em Nature} {\bf 409}, 641--645 (2001).

\bibitem{Ref_62J_from_PRE_2004}
Y. Wei, S. Kim, D. Fela, J. Baum and M. H. Hecht,
``Solution structure of a de novo protein from a designed combinatorial library'',
{\em Proc. Natl. Acad. Sci. USA} {\bf 100}, 13270--13273 (2003).




\bibitem{Ref_44_from_PRE_2004}
D. Baker,
``A surprising simplicity to protein folding'',
{\em Nature} {\bf 405}, 39--42 (2000).

\bibitem{Ref_47_from_PRE_2004}
F. Ding, N.~V. Dokholyan, S.~V. Buldyrev, H.~E. Stanley and E. I. Shakhnovich,
``Direct Molecular Dynamics Observation of Protein Folding Transition State Ensemble'',
{\em Biophys. J.} {\bf 83}  3525--3532 (2002).



\bibitem{Ref_64A_from_PRE_2004}
J.~U. Bowie, J.~F. Reidhaar-Olson, W.~A. Lim and R. T. Sauer,
``Deciphering the message in protein sequences: tolerance to amino acid substitutions'',
{\em Science} {\bf 247}, 1306--1310 (1990).

\bibitem{Ref_64B_from_PRE_2004}
W.~A. Lim and R.~T. Sauer,
``The role of internal packing interactions in determining the structure and stability of a protein'',
{\em J. Mol. Biol.} {\bf 219}, 359--376 (1991).

\bibitem{Ref_64C_from_PRE_2004}
D.~W. Heinz, W.~A. Baase, and B.~W. Matthews,
``Folding and function of a T4 lysozyme containing 10 consecutive alanines illustrate the redundancy of information
in an amino acid sequence'',
{\em Proc. Natl. Acad. Sci. USA} {\bf 89}, 3751--3755 (1992).

\bibitem{Ref_64D_from_PRE_2004}
B.~W. Matthews, ``Structural and genetic analysis of protein stability'',
{\em Annu. Rev. Biochem.} {\bf 62}, 139--160 (1993).



\bibitem{Ref_66_from_PRE_2004}
L. Holm and C. Sander,
``An evolutionary treasure: unification of a broad set of amidohydrolases related to urease'',
{\em Proteins: Structure, Function and Bioinformatics} {\bf 28}, 72--82 (1997).

\bibitem{Ref_87_from_PRE_2004}
N. Nagano, C.~A. Orengo, and J.~M. Thornton,
``One fold with many functions: the evolutionary relationships between TIM
barrel families based on their sequences, structures and functions'',
{\em J. Mol. Biol.} {\bf 321}, 741--765 (2002).


\bibitem{franklin_elife_2018}
M. W. Franklin, S. Nepomnyachyi, R. Feehan, N. Ben-Tal, R. Kolodny and J.~S.~G. Slusky,
``Evolutionary pathways of repeat protein topology in bacterial outer membrane proteins'',
{\em eLife} {\bf 7}, e40308 (2018).


\bibitem{Ref_67_from_PRE_2004}
D.~T. Jones, W.~R.Taylor and J.~M. Thornton,
``A new approach to protein fold recognition'',
{\em Nature} {\bf 358}, 86--89 (1992).



\bibitem{Ref_36_from_PRE_2004}
E.~V. Koonin, Y.~I. Wolf and G.~P. Karev,
``The structure of protein universe and genome evolution'',
{\em Nature} {\bf 420} 218--223 (2002).

\bibitem{Ref_79_from_PRE_2004}
G. Tiana, B.~E. Shakhnovich, N.~V. Dokholyan and E.~I. Shakhnovich,
``Imprint of evolution on protein structures'',
{\em Proc. Natl. Acad. Sci. USA} {\bf 101} 2846--2851 (2004).


\bibitem{bernal}
J. D. Bernal,
``Structure of proteins'',
{\it Nature} {\bf 143}, 663--667 (1939).

\end{thebibliography}
\end{document}